\documentclass{article}

\newcommand{\Oh}[1]
  {\ensuremath{\mathcal{O}\!\left( {#1} \right)}}
\newcommand{\occ}
  {\ensuremath{\mathit{occ}}}

\begin{document}

\title{Searching and Indexing Genomic Databases\\via Kernelization}
\author{Travis Gagie and Simon J.\ Puglisi}
\date{\today}
\maketitle

\begin{abstract}
The rapid advance of DNA sequencing technologies has yielded databases of thousands of genomes.  To search and index these databases effectively, it is important that we take advantage of the similarity between those genomes.  Several authors have recently suggested searching or indexing only one reference genome and the parts of the other genomes where they differ.  In this paper we survey the twenty-year history of this idea and discuss its relation to kernelization in parameterized complexity.
\end{abstract}

\section{Introduction}
\label{sec:introduction}

The Human Genome Project took thirteen years and three billion dollars to sequence a human genome, but the latest next-generation sequencing methods take only a few days and a few thousand dollars.  With these methods, initiatives such as the 1000 Genomes Project and the 100\,000 Genomes Project are now feasible.  Advances in sequencing have far outstripped advances in computer processors and random-access memory, however, so it is increasingly challenging to make use of the data available.  For example, while modern aligners can easily hold in memory the index for approximate pattern matching on a single human genome, they cannot handle thousands of human genomes.  Schneeberger et al.~\cite{SHOWGKW09} proposed that we index the common parts of the genomes only once for them all, but we index the parts near variation sites for each genome.  Ferrada~\cite{FGHP14} suggested indexing the parts of all the genomes near boundaries between phrases in the LZ77 parse of the database.  This is more general and may give better compression but requires the LZ77 parse, which is difficult to compute when the database does not fit in memory.  Wandelt et al.~\cite{WSBL13} proposed using a modified parse in which phrases must occur in a reference genome, which is easier to compute.  (When papers have appeared in journals we cite those versions, although their chronological order may differ from that of previous versions.)  Danek et al.~\cite{DDG14} recently showed that with this general approach we can store an index for approximate pattern matching on the database from the 1000 Genomes Project, in the memory of a commodity personal computer.  This has so far not been possible with competing approaches, as surveyed by Vyverman et al.~\cite{VDFD12}.

When we are not given an upper bound on the pattern length, we can use one of the competing indexes that does not require such a bound or we can scan, with an online pattern-matching algorithm, the reference genome and the parts of the other genomes near phrase boundaries.  Wandelt and Leser~\cite{WL12} and Rahn et al.~\cite{RWR14} proposed the latter idea specifically for approximate pattern matching in genomic databases, but the general approach has a twenty-year history in the field of compressed pattern matching.  In this paper we survey that history and relate it to current research: in Section~\ref{sec:compression} we discuss some relevant data compression schemes and how they have been augmented to support fast random-access reading; in Section~\ref{sec:searching} we discuss how they have been used to speed up pattern-matching; in Section~\ref{sec:indexing} we discuss how they have been used in compressed indexing.  While writing this survey, we realized that scanning or indexing only parts of the database and then mapping the solution for those parts onto a solution for the whole database, is like kernelization in parameterized complexity.  (We note that kernels in parameterized complexity bear no relation to operating system kernels nor to kernels in machine learning.)  We emphasize this perspective because we feel that computing a pattern-matching kernel is an interesting problem in itself, regardless of how we process it later, and deserving of further study.  Of course, the nature and even the existence of the kernel depend on the problem we are trying to solve.

\section{Compression with Random-Access Reading}
\label{sec:compression}

In general, the best compression of highly repetitive datasets is achieved with the LZ77 algorithm by Ziv and Lempel~\cite{ZL77}.  Suppose \(S [1..n]\) is a string with \(S [n] = \$\), which is an end-of-file symbol that does not occur elsewhere in $S$.  LZ77 works by parsing $S$ into phrases such that, for each phrase \(S [i..j]\), \(S [i..j - 1]\) occurs in \(S [1..j - 2]\) but \(S [i..j]\) does not occur in \(S [1..j - 1]\); that phrase is stored as a triple consisting of a pointer to \(S [i..j]\)'s first occurrence in $S$ (which is called the phrase's source), \(j - i\), and \(S [j]\).  The LZ77 encoding of $S$ takes $\Oh{z \log n}$ bits, where $z$ is the number of phrases in the parse.  For example, in the following verses vertical lines indicate phrase boundaries:
\begin{quotation}
\noindent 9$\vert$9-$\vert$b$\vert$o$\vert$t$\vert$tl$\vert$e$\vert$s$\vert$-o$\vert$f$\vert$-be$\vert$er$\vert$-on$\vert$-t$\vert$h$\vert$e-$\vert$w$\vert$a$\vert$ll$\vert$-9$\vert$9-bottles-of-beer-\\
I$\vert$f-o$\vert$n$\vert$e-o$\vert$f-t$\vert$ho$\vert$se$\vert$-bottles-s$\vert$hou$\vert$ld$\vert$-h$\vert$ap$\vert$pe$\vert$n-to$\vert$-f$\vert$all-\\
98$\vert$-bottles-of-beer-on-the-wall-\\[2ex]
98$\vert$-bottles-of-beer-on-the-wall-98-bottles-of-beer-\\
I$\vert$f-one-of-those-bottles-should-happen-to-fall-\\
97$\vert$-bottles-of-beer-on-the-wall\dots
\end{quotation}
(We have displayed the verses with linebreaks to increase readability, but we have not considered them while computing the parse.)  Although these verses may be annoyingly similar by the standards of natural language, they are far less similar than human genomes.  Indeed, most repetitive biological datasets are much too similar (as well as much too large) for us to use them as informative examples.

One drawback of LZ77 compression is that reading a character in a compressed string can be very slow.  Rytter~\cite{Ryt03} and Charikar et al.~\cite{CLLPPSS05} showed how we can turn that parse into a balanced straight-line program (SLP) for $S$ with $\Oh{z \log n}$ rules.  An SLP for $S$ is a context-free grammar in Chomsky normal form that generates $S$ and only $S$; it is balanced if the height of each subtree in the parse tree is logarithmic in that subtree's size.  It follows from Rytter's and Charikar et al.'s results that we can store $S$ in $\Oh{z \log^2 n}$ bits and support random-access reading of any substring of $S$ with length $\ell$ in $\Oh{\log n + \ell}$ time.  Verbin and Yiu~\cite{VY13} showed that this is nearly optimal in the worst case.  Bille et al.~\cite{BLRSSW11} showed how, given even an unbalanced SLP for $S$ with $r$ rules, we can store $S$ in $\Oh{r \log n}$ bits and support random-access reading in $\Oh{\log n + \ell}$ time.  Rytter's, Charikar et al.'s and Bille et al.'s constructions are not practical, but there are practical grammar-based compressors, such as those by Larsson and Moffat~\cite{LM99} and Maruyama and Tabei~\cite{MT14}.  As far as we know, block graphs by Gagie et al.~\cite{GGP11,GHP14} are the most practical grammar-like representations for random-access reading.  The LZ78 algorithm by Ziv and Lempel~\cite{ZL78} does not compress repetitive datasets as well as LZ77, but the LZ78 encoding of $S$ can easily be augmented to support random-access reading in $\Oh{\log \log n + \ell}$ time.  LZ78 also works by parsing $S$ into phrases but then each phrase must extend a previous phrase plus one character.  Because of this property, the LZ78 encoding of $S$ has \(\Omega (\sqrt{n})\) phrases, even when \(S = a^n\).

In the example above, the first verse contains many phrase boundaries but the second verse contains only three.  Kuruppu et al.~\cite{KPZ10} proposed that, given a set of similar strings (or one string that can easily be divided into similar substrings), we store the first string in plain text as a reference and compress the others with a version of LZ77 that restricts phrases' sources to occur in the reference.  They called this scheme Relative Lempel-Ziv (RLZ) and showed it compresses genomic databases very well in practice (although it too uses \(\Omega (\sqrt{n})\) phrases, even when \(S = a^n\)) and there are several implementations of this approach, such as those by Deorowicz and Grabowski~\cite{DG11}, Kuruppu et al.~\cite{KBCZ12} and Ferrada et al.~\cite{FGGP14}.  Even when there is no obvious reference, Kuruppu et al.~\cite{KPZ11} showed we can often build one by sampling the dataset: intuitively, if a substring is common then it is likely to appear in our sample, and if it is not then we lose little by not compressing it well; this can be formalized using results about SLPs.

\section{Searching}
\label{sec:searching}

Farach and Thorup~\cite{FT98} observed that the first occurrence of any pattern \(P [1..m]\) in $S$ must cross or end at a phrase boundary in the LZ77 parse.  K\"arkk\"ainen and Ukkonen~\cite{KU96} showed how, if we already know the locations of $P$'s occurrences in $S$ that cross or end at phrase boundaries, then we can deduce the locations of all its other occurrences from the structure of the parse.  By the same arguments, LZ78 also has these properties and Karpinski, Rytter and Shinohara~\cite{KRS97} simultaneously proved similar results for SLPs.  Bille et al.~\cite{BFG09} observed that any substring of $S$ within edit distance $k$ of $P$ (i.e., any of $P$'s approximate matches) has length at most \(m + k\), and any such substring that does not cross or end at an LZ78 phrase boundary must be an exact copy of an earlier one that does.  They gave an algorithm for approximate pattern matching in LZ78 strings that works by extracting the \(m + k\) and \(m + k - 1\) characters before and after each LZ78 phrase boundary, respectively, using a technique similar to those discussed in Section~\ref{sec:compression}; scanning the resulting substrings with any online algorithm for approximate pattern matching in uncompressed strings; and then deducing the locations of the other approximate matches from the structure of the parse.

Bille et al.~\cite{BLRSSW11} extended this approach to show how we can find all $P$'s approximate matches in $S$ from an SLP for $S$.  Recently, Gagie et al.~\cite{GGP14} extended it further to show how we can preprocess the LZ77 parse of $S$ in $\Oh{z \log n}$ time such that later, given $P$ and $k$, we can find all $P$'s $\occ$ approximate matches in $\Oh{z \min (m k, m + k^4) + \occ}$ time.  Their algorithm works by extracting the \(m + k\) and \(m + k - 1\) characters before and after each LZ77 phrase boundary, respectively, and then continuing as with the algorithm by Bille et al.~\cite{BFG09}.  The set of substrings we extract is like a kernel in parameterized complexity: the total length of the substrings can be much smaller than $n$, but a solution on them can quickly be mapped to a solution on all of $S$.  For our example from Section~\ref{sec:compression} with \(m = 4\) and \(k = 1\), the kernel is
\begin{quotation}
\noindent 99-bottles-of-beer-on-the-wall-99-bo\\
eer-If-one-of-those-bottles-should-happen-to-fall-98-bot\\
ll-98-bot\\
eer-If-on\\
ll-97-bot\dots\
\end{quotation}
If we want a kernel consisting of only a single string, we can concatenate the substrings with \(k + 1\) copies of $\$$ between each consecutive pair.  Notice that if we are careful, we can avoid scanning the fourth substring ``eer-If-on'', since it occurs in the second substring.

We do not wish to leave the impression that kernelization is the only approach used in compressed pattern matching, nor even that the papers mentioned above are the only ones that use it.  We have focused on those papers because we feel they are the most relevant to the practical bioinformatics papers by Wandelt and Leser~\cite{WL12} and Rahn et al.~\cite{RWR14} mentioned in Section~\ref{sec:introduction}.  Those authors were apparently unaware of the field of compressed pattern matching and re-invented kernelization specifically for approximate pattern matching in genomic databases, with kernels based on RLZ instead of LZ77, LZ78 or SLPs.  This may be because the earlier researchers using kernelization for pattern matching did not publicize their ideas in interdisciplinary forums or implement their ideas in tools usable by other scientists.

\section{Indexing}
\label{sec:indexing}

K\"arkk\"ainen and Ukkonen~\cite{KU96} gave the first LZ-based index, which supported exact pattern matching and stored $S$ separately and uncompressed.  They used Patricia trees and range reporting to find a set of candidate matches crossing or ending at LZ77 phrase boundaries; verified them by checking $S$; and then used more range reporting to find the other matches.  We can obtain various time-space tradeoffs by compressing $S$ and use the methods discussed in Section~\ref{sec:introduction} to extract the characters needed to verify candidate matches.  Claude and Navarro~\cite{CN12} modified K\"arkk\"ainen and Ukkonen's index to use a grammar-compressed encoding of $S$, and Kreft and Navarro~\cite{KN13} modified it to use the encoding of $S$ produced by a version of LZ77 they called LZ-End, which supports fast random-access reads starting at phrase boundaries.  Arroyuelo et al.~\cite{ANS12} and Do et al.~\cite{DJSS14} gave indexes based on LZ78 and RLZ, respectively, and Maruyama et al.~\cite{MNKS13} and Takabatake et al.~\cite{TTS14} gave indexes based on the edit-sensitive parsing by Cormode and Muthukrishnan~\cite{CM07}.  Gagie et al.~\cite{GGKNP14} recently gave a version of K\"arkk\"ainen and Ukkonen's index that uses a total of $\Oh{z \log^2 n}$ bits and returns the locations of all $P$'s $\occ$ occurrences in $S$ in $\Oh{m \log m + \occ \log \log n}$ time.  These indexes require no assumptions about the pattern.

K\"arkk\"ainen and Sutinen~\cite{KS98} gave an index based on a version of LZ77 that allows phrases to overlap by \(q - 1\) characters, where $q$ is a parameter.  If $P$ has length exactly $q$, then their index returns the locations of all $P$'s occurrences in $S$ in optimal $\Oh{m + \occ}$ time.  If we are given an upper bound $M$ on the pattern length at construction time, then even with K\"arkk\"ainen and Ukkonen's original version, we need keep only a kernel of the text and can use $\Oh{z \log n + z M \log \sigma}$ bits in total, where $\sigma$ is the size of the alphabet.  We suspect this escaped investigation for so long because it seemed too obvious and inelegant to be theoretically interesting, and the need to index massive, highly repetitive datasets in practice has become pressing only since the development of next-generation sequencing methods.

The use of kernelization for indexing was eventually investigated by Schneeberger et al.~\cite{SHOWGKW09}, although they did not present kernelization as a separate process because their work was application-driven.  As noted in Section~\ref{sec:introduction}, they proposed that, given a database of genomes from the same species, we index the common parts of the genomes only once for them all, but we index the parts near variation sites for each genome.  Wandelt et al.~\cite{WSBL13} and Danek et al.~\cite{DDG14} gave similar results, essentially using a kernel based on the RLZ parse.  Like Schneeberger et al., these authors indexed the kernels using specific methods based on q-grams or seeds.  Danek et al.'s index for the database for the 1000 Genomes Project is the first one to fit in the memory of a commodity personal computer.  Ferrada et al.~\cite{FGHP14} emphasized kernelization (albeit not under that name) in terms of the LZ77 parse, which is more general and may give better compression, and pointed out that we can use any index for approximate pattern matching to store the kernel.  One point they did not comment on, and which we hope to have clarified in this paper, is that we can consider kernels based on LZ77, LZ78, RLZ, other compression schemes, or possibly other algorithms entirely.  These kernels may be easier to compute when the database does not fit in memory, or have other useful properties that make them preferable in some situations.  One interesting problem is how we can best maintain a dynamic kernel for an expanding database.  This could allow us to align reads against a genomic database and then add the newly-assembled genome, which could be useful when dealing with mutating cancer genomes or changing strains of a disease during an outbreak.

\section*{Acknowledgement}

Many thanks to Fabio Cunial, Pawe\l\ Gawrychowski, Simon Grabowski, Juha K\"arkk\"ainen, Veli M\"akinen, Gonzalo Navarro, Esa Pitk\"anen, Yasuo Tabei and Niko V\"alim\"aki, for helpful discussions.

\bibliographystyle{plain}
\bibliography{frontiers}

\end{document}